\documentclass[twocolumn,showpacs,10pt]{revtex4}
\usepackage{graphicx}
\usepackage{bm}
\begin{document}

\title{$^3PF_2$ pairing in high-density neutron matter}

\author{J. M. Dong $^{1}$, U. Lombardo $^{2}$, and W. Zuo $^{1}$}
\affiliation{\mbox{$^{1}$ Institute of Modern Physics, Chinese
Academy of Sciences, Lanzhou 730000, China}\\
\mbox{$^{2}$ Universita' di Catania and Laboratori Nazionali del Sud
(INFN), Catania 95123, Italy}}
\date{ \today}


\begin{abstract}
The onset of the $^3PF_2$ superfluid phase in high-density neutron
matter is studied within the BCS framework with two and three body
forces. Owing to the strong correlations the energy gap is so sizeably quenched
as to demand to reconsider the role of superfluidity in neutron-star phenomena.
\end{abstract}

\maketitle

\section{Introduction}

So far, the neutron stars (NS) have been considered as a
rich laboratory of various superfluid phases of
nuclear matter \cite{FT1,FT2,FT3,FT4,FT5,FT6,FT7,FT8,FT9,FT10}.
Recently the interest has been focussed on the NS
interior, where both the vortex pinning
responsible for the observed period glitches\cite{pizzo} and  the nucleon
superfluidity responsible for the main cooling mechanisms
\cite{cassi}  are supposed to be located. In particular the recent
observations of cooling in the NS of Cassiopeia A have been
considered to be a direct evidence of the $^3PF_2$ pairing in NS
core, and the energy gap needed to explain the data has been estimated to be around
$0.1$ MeV\cite{cassi}. From the theoretical
viewpoint, the onset of the $^3PF_2$ superfluid state occurs in
regions above the saturation density. This event rises the
problem of the pairing formation in high density  nuclear matter, that however all previous calculations neglect (see Ref.\cite{schul} and references therein).

The density region, where the $^3PF_2$ component of the nuclear
interaction is the most attractive, may extend to several times the
saturation density of nuclear matter. In such a high-density regime,
the short-range correlations are so strong that the momentum
distribution around the Fermi level departs significantly from the
typical profile of a degenerate Fermi system. This departure is
measured by the so-called $Z$-factor $(0<Z<1)$ \cite{migdal}. Since
the deformation of the Fermi surface hinders particle transitions
around the Fermi energy $\epsilon_F$, the pairing gap is expected to
get reduced in any case, even for the low density $^1S_0$ pairing
channel (see,e.g.,\cite{cao}). Concerning the high-density
neutron-neutron pairing in the $^3PF_2$ channel, the calculations so
far reported in the literature ignore the effect of the $Z$-factor,
owing to the large uncertainty still existing in the magnitude of
the nucleon-nucleon ($NN$) interaction in the $^3PF_2$ channel,
including both the two body force (2BF) and the three-body force (3BF) component.
Other many-body effects, such
as screening effect, have been neglected for the same reason.

In this note, we present a study of the $Z$-factor effect on the
$^3PF_2$ pairing in pure neutron matter. In principle, we should
consider asymmetric nuclear matter for application to NS core, but
the small proton fraction is not relevant in this
context, as discussed below. The deformation of the Fermi surface
and the $Z$-factor are studied in the framework of the Brueckner
theory with 2BF and 3BF \cite{zhl}. The energy gap is then
calculated within the generalized BCS theory \cite{zuo},
including in the  pairing interaction not only 2BF, but also 3BF. The latter, in fact, is dominant at high density and therefore it is expected to directly  influence  pairing gap in addition to the Z-factor.
\section{Formalism and results}
\subsection{Nucleon propagator in neutron matter}
The neutron Green's function is given by
\begin{equation}
G^{-1}(p,\omega) = \omega - \frac{p^2}{2m} -
\Sigma(p,\omega) + e_F,
\end{equation}
where $e_F$ denotes the Fermi energy and 
$\Sigma(p,\omega)$ the self-energy. Expanding the latter in a series of powers of
the quasiparticle energy around the Fermi surface, we obtain
\begin{equation}
G^{-1}(p,\omega) \approx
Z(p)^{-1}(\omega - \epsilon_p),
\end{equation}
where the quasi-particle energy and the quasi-particle strength are
\begin{eqnarray}
\nonumber
\epsilon_p = \frac{p^2}{2m}+\Sigma(p,\epsilon_p)-e_F \\
\nonumber
Z(p) = \left[ 1-\frac{\partial\Sigma(p,\omega)}{\partial\omega}\right]_{\omega=\epsilon_p}^{-1},
\end{eqnarray}
respectively. The quasiparticle strength $Z(p)$
measures the deviation of a correlated Fermi system from the ideal
degenerate Fermi gas. The occupation numbers  $n(p)$ and the $Z(p)$ factors have been calculated in the framework of the Brueckner theory \cite{day,jeukenne}. Grace to the inclusion of the three body forces  the hole-line expansion can be extended up to high densities.
The self-energy, truncated to the second order (see Ref. \cite{ebhf} for details), provides us with  a good reproduction of the empirical  nuclear  mean field \cite{mah} and the optical-model potential \cite{omp}, so that we are quite confident that the next orders  are irrelevant for the present calculation.
 We employ as 2BF the meson exchange  Bonn B potential \cite{machleidt}, whose meson parameters are constrained by the fit with the experimental phase shifts of NN scattering. The microscopic meson exchange 3BF is the one constructed by Li {\it et al.} \cite{zhl}. It is consistent with the 2BF because it adopts the same meson  parameters as Bonn B, so that there are no free parameters in the model.

The $Z(p)$ factor is related to the depletion of the
occupation number $n(p)$ around the Fermi surface.
According to the Migdal-Luttinger theorem \cite{ml}
its value $Z=Z(p^F)$(Z-factor) equals the discontinuity of the occupation number
at the Fermi surface, i.e.,
\begin{equation}
\lim_{\varepsilon\rightarrow 0} [
n(p^F-\varepsilon)-n(p^F+\varepsilon)] = Z(p^F) ,
\end{equation}
where $p^F$ is the Fermi momentum.
In our approximation $\Sigma(p,\omega) =\Sigma_1(p,\omega)+\Sigma_2(p,\omega)$,
where $\Sigma_1(p,\omega)$ determines the left limit and $\Sigma_2(p,\omega)$ the right limit of
the preceding equation for $\varepsilon \rightarrow 0$.
\begin{figure}[htb]
\includegraphics[scale=0.6]{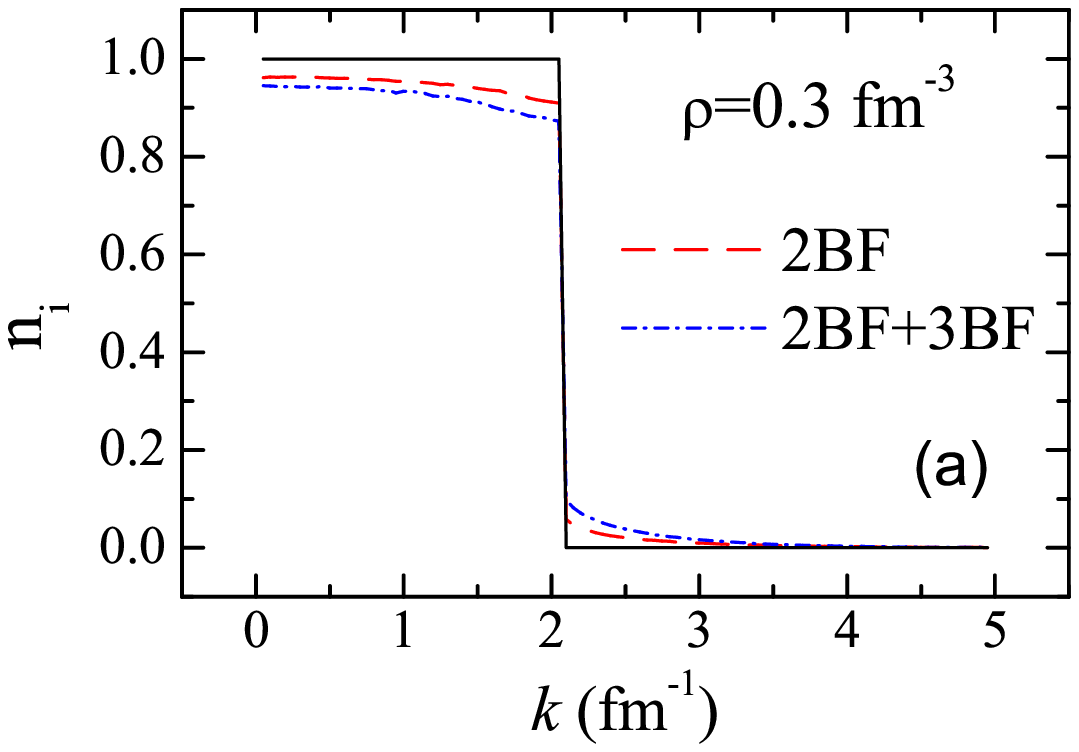}
\end{figure}

\begin{figure}[htb]
\includegraphics[scale=0.6]{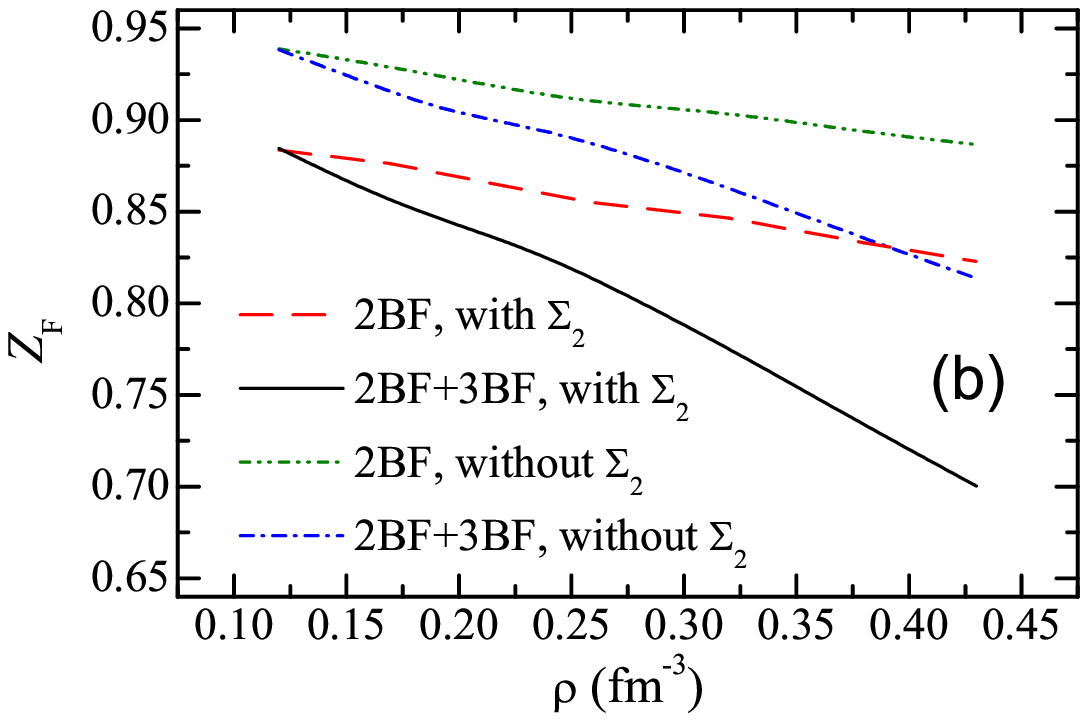}
\caption{(Color Online) Occupation numbers
vs. momentum at a given density (upper panel) and $Z$-factors  vs. density (lower panel) in pure neutron matter. The effect of 3BF is shown in both panels. In the lower panel the calculations are reported for two approximations of the self-energy}
\end{figure}

 In Fig. 1(a), we display the calculated occupation probability in pure neutron matter at density
 $\rho=0.3$ fm$^{-3}$. One easily observes the remarkable deviation from the ideal Fermi gas (solid line) due to
 the strong short-range correlations. As expected, the deviation is slightly enhanced by 3BF.
In Fig. 1(b) the calculated $Z$-factor is displayed
vs.density in the two different approximations for the self-energy,
i.e., $\Sigma\approx\Sigma_1$ and $\Sigma\approx\Sigma_1+\Sigma_2$,
respectively. The calculation of $Z_F$ from Eq.(3) requires a high numerical
accuracy: increasing the accuracy the calculated $Z_F$ gets lower and lower
until the converging value is reached.
It is noticed that without 3BF, the
$Z$-factors decrease slowly as a function of density.
Adding the contribution of $\Sigma_2$ leads to an overall
reduction of the $Z$ factor. The 3BF reduces further the
$Z$-factor and its effect increases rapidly with density.
As a consequence, including 3BF makes the decrease
of the $Z$-factor as a function of density much rapid than that
obtained by adopting pure 2BF. Therefore
 3BF  induces a strong extra deviation from the ideal
 Fermi gas model.

\subsection{Gap equation in the  $^3PF_2$ channel}

The $^3PF_2$ superfluidity in pure neutron matter has been
investigated by using various theoretical approaches
\cite{TB1,TB2,TB3,TB4,TB5,TB6,TB7} with 2BF, and later extended to
 microscopic 3BF forces by Zuo {\it et al.} \cite{zuo}.
In this case, the pairing gaps are determined by the two coupled
equations:
\begin{eqnarray}
\left( \begin{array}{l} \Delta_L ( p)\\ \Delta_{L+2}( p) \end{array} \right)
 = -\frac{1}{\pi}\int_0^\infty \!  p'^2{\rm d} p'  \frac{Z(p) Z(p')}{E_{p'}}\hspace*{1.5cm}\nonumber\\
\times \left( \begin{array}{ll} V_{L,L}(p,p') & V_{L,L+2}(p,p') \\ V_{L+2,L} (p,p') &
  V_{L+2,L+2}(p,p')
 \end{array} \right)
 \left(\begin{array}{l} \Delta_L  (p') \\ \Delta_{L+2} (p')\end{array} \right),\ \
\end{eqnarray}
where $E_{p}^2 = (\epsilon_p-\mu)^2 + \Delta_p^2 $ and
$\Delta^2=\Delta_L^2+\Delta_{L+2}^2$.  $V_{L,L'}(p,p')$ are the
matrix elements of the realistic $NN$ interaction in the coupled
$^3PF_2$ channel.
In the gap equation, the $Z$-factors and the single-particle energy $\epsilon_p$
are calculated from the Brueckner theory.
As for the pairing interaction $V_{L,L'}(p,p')$ in the $^3PF_2$ coupled channel,
we adopt the same 2BF and 3BF as in the Brueckner calculation.
The 3BF cannot be neglected in the gap calculation because the $^3PF_2$
pairing is expected to occur in the high density domain, where
3BF is quite sizeable specially in the $^3PF_2$ channel.

The results are summarized in Fig.2. Neglecting the Z-factor effect
(upper panel), the magnitude of the $^3PF_2$ gap with the new
interaction dos not differ from previous calculations
with AV18 potential \cite{zuo}. On the other hand, the 3BF enhances
the $^3PF_2$ superfluidity significantly at higher
densities. But,as shown in the lower
panel of Fig.2, the introduction of the $Z$-factor  effect
fatally quenches the pairing gaps to a value less  0.05 MeV, one order of magnitude
smaller than the value with full interaction.
The effect of the $Z$-factor appears to be extremely sizeable at high densities.
It is worth noticing that this effect is opposite to the 3BF effect on
the $^3PF_2$ pairing in neutron matter: the former turns out to
be much stronger than the latter. In conclusion, the departure of the system from the pure degenerate limit drives the pairing attenuation at high density, regardless of whether the 3BF is included or not.

\begin{figure}[htb]
\includegraphics[scale=0.8]{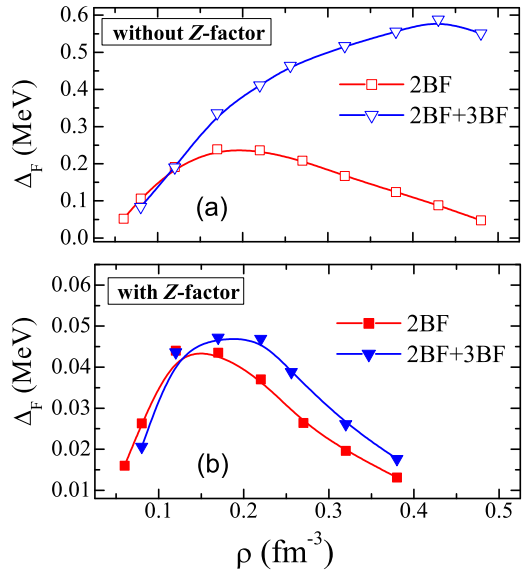}
\caption{(Color Online)Effect of the Fermi surface depletion on $^3PF_2$ pairing gap in pure neutron matter. Notice the y-scale change from (a) to (b)}
\end{figure}

In Ref. \cite{lombardo:2001}, the effect of the $Z$-factor was investigated also for the $^1S_0$ pairing
in neutron matter, and it was shown to reduce the energy gap.
Comparing with the results of Ref.\cite{lombardo:2001}, the same effect
on the $^3PF_2$ superfluidity appears to be much stronger. This is simply due to the fact that
in the density region of the $^3PF_2$ superfluidity
the deviation of neutron matter from the degenerate Fermi gas limit becomes much larger.

\section{Conclusions}

In conclusion, we have studied the anisotropic $^3PF_2$ pairing in
pure neutron matter. The effects of the Fermi surface depletion (Z-factor) have been included in the calculation of the energy gap. In the pure degenerate limit, the $^3PF_2$ superfluid phase extends over a broad density range with a gap  peak value of about 0.2 MeV without 3BF and 0.5 MeV with 3BF. The inclusion of the $Z$-factor leads to a rapid decrease of the gap magnitude by one order of magnitude: its peak value
drops  to less than 0.05 MeV and the superfluidity domain shrinks to $0.1-0.4$ fm$^{-3}$.
In neutron stars the proton fraction in
$\beta$-equilibrium with neutrons could in principle affect the
$^3PF_2$ pairing, but in the non-vanishing gap range  it is less
than $15\%$ of the total density. Recent calculations on the isospin
dependence of the quasi-particle strength show that the enhancement
of the neutron $Z$-factor is negligible for such small proton fraction
\cite{zfac}. Even including the additional screening suppression
\cite{cao}, the conclusion remains that the departure  from the
Fermi gas limit is the main cause of pairing disappearance in high-density nuclear matter.
This  result makes  doubtful  the role of the $^3PF_2$  pairing
in  NS core.

\section{Acknowledgments}

Discussions with D. Yakovlev are gratefully acknowledged. This work
was supported by the Major State Basic Research Developing Program
of China under No. 2013CB834405, the National Natural Science
Foundation of China under Grants No. 11175219, 10975190, 11275271
and 11205075; the Knowledge Innovation Project (KJCX2-EW-N01) of
Chinese Academy of Sciences, and the Funds for
Creative Research Groups of China under Grant No. 11021504.

\end{document}